\documentclass[12pt]{iopart}
\usepackage{iopams} 
\usepackage{graphicx}
\usepackage[FIGTOPCAP, raggedright, nooneline]{subfigure}
\expandafter\let\csname equation*\endcsname\relax
\expandafter\let\csname endequation*\endcsname\relax
\usepackage{amsmath}
\usepackage{latexsym}
\usepackage{multirow}
\usepackage{xcolor}
\usepackage[separate-uncertainty,range-phrase = --]{siunitx}
\usepackage{nicefrac}
\usepackage{tikz}
\usepackage{cite}

\usepackage[normalem]{ulem}
\graphicspath{{../v01/figure_files/}{../v01/}}

\begin{document}
	
	\title{Instability and fingering of interfaces in growing tissue}
	
	\author{Tobias B\"uscher$^{1}$, Angel L. Diez$^{1,2}$, Gerhard Gompper$^1$ and Jens Elgeti$^{1,*}$}
		\address{$^1$Theoretical Soft Matter and Biophysics, Institute of
			Complex Systems and Institute for Advanced Simulation, Forschungszentrum J\"ulich, 52425 J\"ulich, Germany}
	\address{$^2$Departamento de Estructura de la Materia, Fisica Termica y Electronica, Facultad de Ciencias Fisicas, Universidad Complutense de Madrid, 28040 Madrid, Spain}
	\ead{j.elgeti@fz-juelich.de}
	
	\date{\today}
	
	\begin{abstract}
Interfaces in tissues are ubiquitous, both between tissue and environment as well as between populations of different cell types. 
The propagation of an interface can be driven mechanically. 
Computer simulations of growing tissues are employed to study the stability of the interface between two tissues on a substrate.
From a mechanical perspective, the dynamics and stability of this system is controlled mainly by four parameters of the respective tissues: (i) the homeostatic stress (ii) cell motility (iii) tissue viscosity and (iv) substrate friction.
For propagation driven by a difference in homeostatic stress, the interface is stable for tissue-specific 
substrate friction even for very large differences of homeostatic stress; however, it becomes unstable above a critical stress difference when the tissue with the larger homeostatic stress has  a higher viscosity. A small difference in directed bulk motility between the two tissues suffices to result in propagation with a stable interface, even for otherwise identical tissues. Larger differences in motility force, however, result in a finite-wavelength instability of the interface. Interestingly, the instability is apparently bound by nonlinear effects and the amplitude of the interface undulations only grows to a finite value in time. 
	\end{abstract}
	
	\maketitle

\section{Introduction}

Interfaces of tissues, their propagation as well as their stability, play an important role in various biological contexts, ranging from tissue development \cite{Batlle2012} to wound healing \cite{Basan2013,  Ravasio2015} and cancer \cite{Bru2003}. In many of these processes, the interface propagates, driven by cell proliferation and/or motility. This leads to the question how the tissue maintains a stable interface, as this is crucial e.g. in development in order to arrive at the desired distinct cell populations, while interface instabilities can have severe consequences, as in cancer metastasis. Several mechanisms act simultaneously in this problem, where each of them can either have a stabilizing or destabilizing effect on the interface. Interfacial tension, e.g. caused by differential adhesion between cell populations \cite{Foty2005}, stabilizes an interface, as it penalyzes increase of interface area. On the other hand, increase of interfacial area can be further amplified, e.g. due to enhanced growth rates in the protruding region, where cells have more free space and access to nutrients, as commonly observed during wound healing \cite{Christini2005, Basan2013, Poujade2007} .

Interface instabilities in systems far from equilibrium are well known in solid-state physics, where several instability mechanisms have been found and studied \cite{Gallaire2017}. Examples are the Saffman-Taylor instability (also known as viscous fingering), which occurs during the injection of a low-viscosity fluid into one of a larger viscosity, the Mullins-Sekerka instability in unidirectional solidification, which arises from the unstable diffusive transport of the latent heat of solidification, and leads to dendritic growth at later stages, and the Rayleigh-Taylor instability between two immiscible fluids when the fluid with higher density is placed on top of the lighter one. Also, in vapor deposition flat interfaces are unstable to roughening, in which the interface width initially grows slowly, but monotonically with time and saturates at a finite value at late times. For tissues, or bacterial colonies as a related example, growth and division of cells can give rise to new instability mechanisms, which, however, may arise from similar mechanisms as the "classical" instabilities of solid-state physics.  For example,  an undulation instability of an incompressible epithelium adjacent to a viscoelastic stroma has been found, where the instability is driven by enhanced growth in the protruding region, which creates a shear flow that builds up pressure at the bottom of the protrusion \cite{Basan2011c}. Coupling cell growth to nutrient diffusion leads to an additional instability, as cells in the protruding region have access to more nutrients, reminiscent of the Mullins-Sekerka instability \cite{Risler2013}. In growing bacterial colonies of \textit{E.~coli} inside a microfluidic device, a streaming instability 
has been oberseved due to steric interactions between large, slow-moving and small, fast-moving cells \cite{Mather2010}. During growth of bacterial colonies on a petri dish, instabilities of the advancing front arise, displaying different levels of complexity, which range from a small number of fingers to densely-branched, dentritic structures \cite{Santalla2018, Giverso2015, Kawasaki1997, Matsushita1990}.

Mechanically regulated propagation of tissues has been studied by employing the concept of homeostatic stress \cite{Ranft2014, Podewitz2016, Williamson2018}. The homeostatic stress is defined as the stress a tisssue exerts onto its surrounding at the state when apoptosis and division balance each other. It has been proposed that in a competition for space between two tissues, the tissue with the lower homeostatic stress (higher homeostatic pressure) grows at the expense of the other \cite{Basan2009}.
 Furthermore, motility forces generated by cells migrating on a substrate can generate stresses on neighboring tissues and affect the competition \cite{Williamson2018}. This has recently been studied by in vitro experiments. Two different confluent cell-layers were initially seperated by a fixed gap. Upon release, the two tissues migrate towards each other and collide ”head on”. 
Interestingly, Ras-transformed Madin-Darby Canine Kidney (MDCK) cells where pushed back by the corresponding wild type cells \cite{Porazinski2016}, while conversely or Ras-transformed Human Embryonic Kidney (HEK) cells outcompeted the corresponding wild type. The cell population which generates larger collective stresses displaces the other population and drives the propagation of the interface between them \cite{Moitrier2019}.

The stability of a propagating interface, driven by homeostatic stress and/or bulk motility differences, between two competing tissues on a substrate has recently been studied by a linear stability analysis \cite{Williamson2018}. Three instability criteria are obtained, where two yield a critical homeostatic stress difference and one a critical difference in motility-force strength above which the interface becomes unstable.

Using a particle-based model of growing tissues \cite{Podewitz2016}, we study the mechanically-regulated competition of two tissues and explore the stability of the interface. Our simulations suggest that nonlinearities provide a strong stabilizing effect on the interface.
Contrary to linear-stability analysis, we find a stable interface when the two tissues differ in their respective substrate friction, even for large homeostatic stress differences. On the other hand, for a different viscosity of the two tissues, an instability arises above a critical difference in homeostatic stress.
However, the instability does not grow forever; instead, a finger-like protrusion of the weaker tissue is left behind in the stronger tissue, which otherwise advances with a broad front.
For a difference in motility-force strength, we find that a low motility has a stabilizing effect onto the interface, causing a decrease of the interface saturation width with growing difference in motility force, while large motility forces cause an unstable interface above a critical point.  Beyond the instability, distinct modes grow strongly in amplitude, but saturate at finite values depending on the strength of motile forces. Hence, the instability due to motility forces seems to be bound by nonlinearities.

Our results demonstrate that the structure of the interface between two competing tissues may serve as a key observable in characterizing mechanical properties of the competing tissues. Indeed, it is often the interfacial properties that reveal malignancy in tumor biology \cite{Tsai2008, Hamdoon2012}.

\section{Simulation model}
Several models have been developed  in order to study mechanical properties and growth of cell monolayer in general and interfaces between different cell types in particular \cite{Byrne2009, VanLiedekerke2015, Block2007}. For example, vertex-based models are commonly employed , e.g. to study physical properties such as shear and compression modulus, or jamming transitions \cite{Alt2017, Staple2010, Murisic2015, Bi2015}. 
We employ the well-established particle-based growth model of Refs. \cite{Marel2014, Podewitz2016}, which has been mapped onto various systems of  growing cell sheets, such as wound healing assays or growth of bacterial colonies in microfluidic devices \cite{Basan2013, Hornung2018}.
A cell is represented by two particles which repel each other via an active growth force 
\begin{equation}
\textbf{\textit{F}}_{ij}^{\text{G}} =
\frac{G}{(r_{ij} + r_0)^2}\hat{\textbf{\textit{r}}}_{ij}\text{,}\label{growth_force}
\end{equation}
with growth-force strength $G$, unit vector $\hat{\textbf{\textit{r}}}_{ij}$ and distance $r_{ij}$ between the two particles and a constant $r_0$.
When the distance between the particles exceeds a threshold $r_{\text{ct}}$ the cell divides. A new particle is then placed in close vicinity of each particle of the mother cell. These pairs constitute the two daughter cells. Particles between different cells interact via a soft repulsive force $\textbf{\textit{F}}_{ij}^{\text{V}}$ on short distances and a constant attractive force $\textbf{\textit{F}}_{ij}^{\text{A}}$ on intermediate distances, where
\begin{equation}
\left.
\begin{array}{@{}ll@{}}
\textbf{\textit{F}}_{ij}^{\text{V}} &= f_0\left(\frac{R_{\text{PP}}^5}{r_{ij}^5}-1\right)\hat{\textbf{\textit{r}}}_{ij} \label{volume}\\
\textbf{\textit{F}}_{ij}^{\text{A}}& = -f_1\hat{\textbf{\textit{r}}}_{ij} 

\end{array}\right\} \text{for } r_{ij}<R_{\text{PP}} \text{,}
\end{equation}
with volume exclusion coefficient $f_0$, adhesion strength $f_1$ and cut-off length $R_{\text{PP}}$. We model apoptosis by removing cells randomly at a constanst rate $k_{\text{a}}$.
Interactions with the underlying substrate are given by a friction force
\begin{equation}
\textbf{\textit{F}}_{i}^{\text{B}}=-\gamma_{\text{b}}\textbf{\textit{v}}_i\text{,}
\end{equation}
with velocity $\textbf{\textit{v}}_i$.
Forces in migrating cell monolayers do not solely arise at the front, but collectively over the whole monolayer \cite{Trepat2009, Moitrier2019}. 
In a simplifyed picture, this is  modeled by a homogeneous bulk motility force \cite{Williamson2018}, given by a constant force perpendicular to the interface
\begin{equation}
\textbf{\textit{F}}_i^{\text{M}}=f_{\text{m}}\cdot\hat{\textbf{\textit{e}}}_x\text{,}
\end{equation}
with motility-force strength $f_{\text{m}}$ and direction $\hat{\textbf{\textit{e}}}_x$ perpendicular to the interface.  This choice of motility model further facilitates comparison of results with Ref.~\cite{Williamson2018}. 
A dissipative particle dynamics thermostat is employed in order to account for energy dissipation and random fluctuation, satisfying the fluctuation-dissipation theorem.
Its temperatute $T$ is chosen low enough that cells can escape local minima, but other thermal effects are small.
Each parameter can be set independently for each cell type and between cell types for inter-cell interactions. 

We define a set of standard-tissue parameters and report simulation parameter relative to these standard values, denoted with a dagger, e.g. $G^\dagger = G/G_0$ (see Tab.~S1 in the SI for numerical values). Time is measured in terms of the inverse apoptosis rate $k_{\text{a}}$ of the standard tissue, distance in terms of the cut-off length $R_{\text{PP}}$ and force in units of $G_0 / R_{\text{PP}}^2$. Thus, the length unit corresponds to the cell size, while time is measured in generations. After one time unit, all cells have divided once on average. Quantities reported in these units are denoted with an asterisk $*$.
We vary the growth-force strength~$G$, the apoptosis rate $k_{\text{a}}$, background friction $\gamma_{\text{b}}$, 
and motility-force strength~$f_{\text{m}}$. The cross-adhesion strength between the two tissues is the same as the adhesion strength within one tissue, as reduced cross-adhesion causes enhanced interfacial growth \cite{Ganai2019}. Thus, no passive interfacial tension is present in our simulations.

We use the "treadmilling simulation setup" introduced in Ref.~\cite{Podewitz2016} in order to obtain steady-state interface progression, by keeping the interface position at the center of the simulation box. All cells are shifted accordingly every 1000 timesteps; excess cells at one end of the simulation box are removed while the weaker tissue replenishes on the other end. In this way, both tissues reach their homeostatic state sufficiently far away from the interface (with system size $L_x^*=140$ in all simulations), thus the interface properties can be studied on long time scales in a computationally efficient way. We measure all quantities in a comoving reference system $s=x-x_0$, with interface position $x_0$. 
\section{Results}

\begin{figure}
	\centering
	\vspace{-0.cm}
	\includegraphics[width=1.\linewidth]{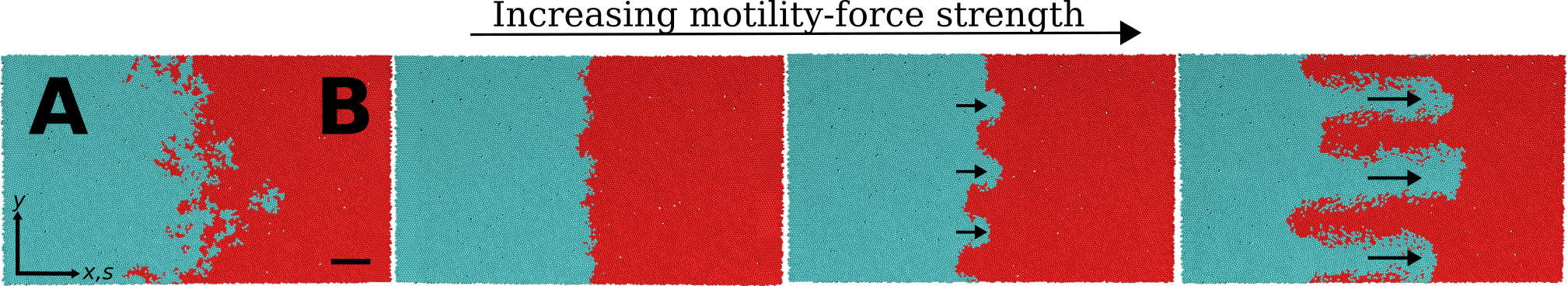}
	\caption{Simulation snapshots in competitions with different motility-force strengts $f_{\text{m}}^{\text{A}\dagger}$ of tissue~A. The tissues are otherwise identical. The interface moves towards tissue~B. From left to right: $f_{\text{m}}^{\text{A}\dagger}=[0,0.002,0.045,0.08]$. The length of the arrows corresponds to the distance the interface moves over one generation (too small for $f_{\text{m}}^{\text{A}\dagger} = 0.002$). System size $L_y^*=80$ and time $t^*=80$ in all. The scale bar is 10 cell sizes. Note that for $f_{\text{m}}^{\text{A}\dagger}\geq0.002$ the snapshots are represantative of the steady state and the undulations do not grow further.}
	\label{snapshots}
\end{figure}

It was shown in Ref.~\cite{Podewitz2016} that the competition between two tissues differing only in homeostatic stress results in a steady-state interface propagation, where the stronger tissue invades the weaker one with a constant velocity. While only stable interfaces were observed in Ref.~\cite{Podewitz2016}, Ref.~\cite{Williamson2018} proposes three different routes to instability for an interface between two competing tissues: (A) For propagation driven by bulk motility, the interface becomes unstable above a critical difference in motility-force strength.  For propagation driven by homeostatic stress, the interface is only unstable under the condition that the two tissues either differ (B) in substrate friction or (C) viscosity. For both cases, (B) and (C), the interface becomes unstable above a case-specific critical difference in homeostatic stress. 
For a combination of both, bulk motility force $f_{\text{m}}^{\text{A/B}}$ and homeostatic stress difference $\Delta\sigma_{\text{H}}=\sigma_{\text{H}}^{\text{B}} - \sigma_{\text{H}}^{\text{A}}$, the interface velocity 

\begin{equation}
v_{\text{int}} = \frac{\Delta\sigma_{\text{H}} + l_{\text{A}}\hat{f}_{\text{m}}^{\text{A}} + l_{\text{B}}\hat{f}_{\text{m}}^{\text{B}}}{ \xi_{\text{A}}l_{\text{A}} + \xi_{\text{B}}l_{\text{B}}} \label{int_velo}
\end{equation}
is predicted, with substrate friction $\xi = 2\gamma_{\text{b}}\rho$, cell density $\rho$, motility-force density $\hat{f}_{\text{m}}^{\text{A/B}} = 2\rho f_{\text{m}}^{\text{A/B}}$, and stress decay length $l=\sqrt{\chi\tau/\xi}$. Here, $\chi$ is the elastic modulus, $\tau$ the time scale at which the tissue loses its elastic character due to cell division and apoptosis, and the product $\chi\tau$ is an effective viscosity. The growth rate $k$ is expanded to linear order around the homeostatic stress as \mbox{$k=\kappa(\sigma-\sigma_{\text{H}})$}, with stress-response coefficient~$\kappa$.  The viscosity is connected to the stress-response coefficient via $\kappa=1/\chi\tau$. For our simulations, these coarse-grained tissue parameters are either direct input parameters, or can be measured in independent single tissue simulations. 

\begin{figure}
	\centering
	\flushleft{\hspace{.cm}(a)\hspace{7.3cm}(b)}\\
	\includegraphics[width=0.49\linewidth]{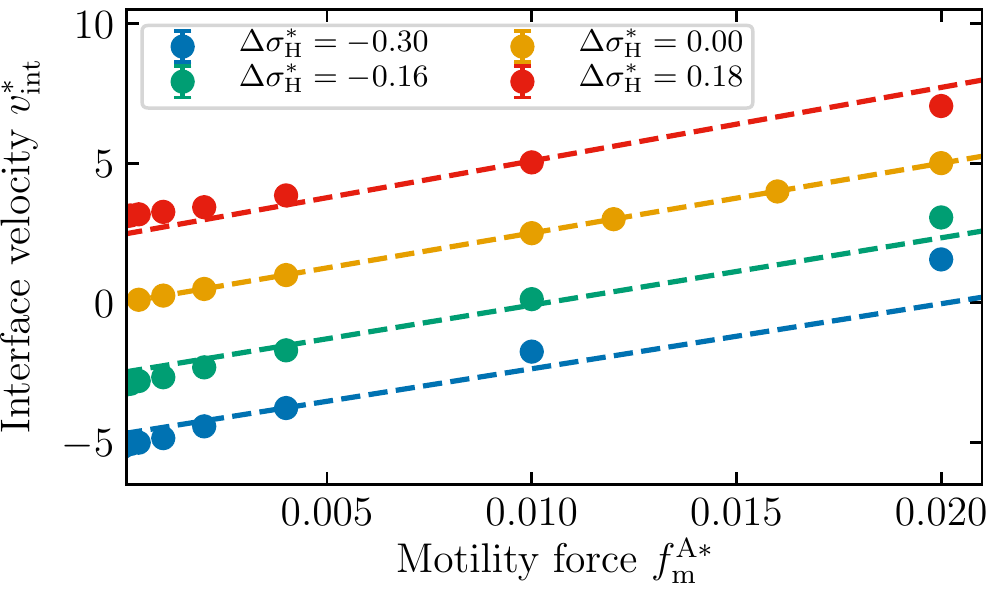}
	\includegraphics[width=0.49\linewidth]{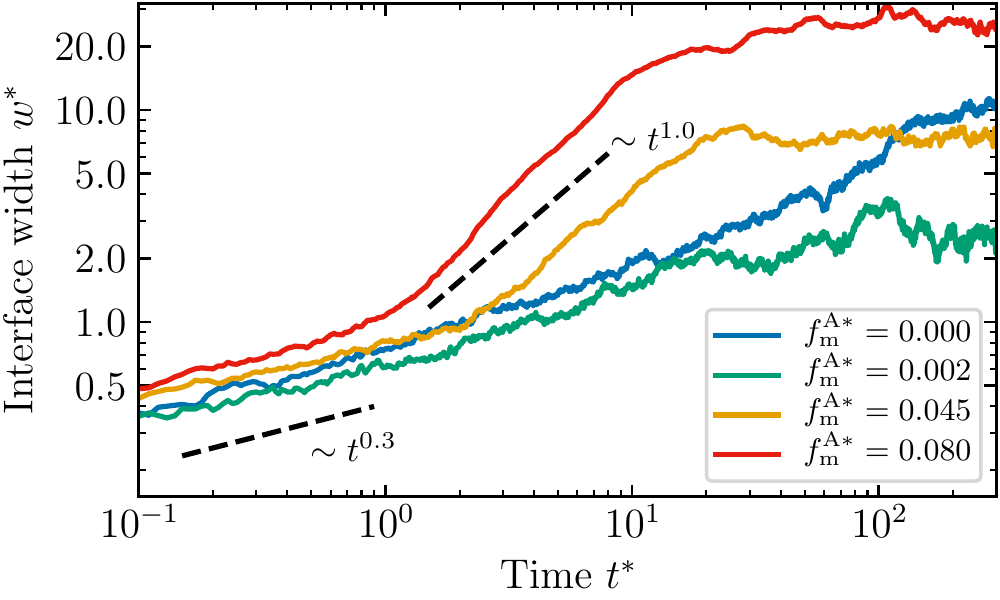}
	\vspace{-.7cm}	\flushleft{\hspace{.cm}(c)\hspace{7.3cm}(d)}\\
	\includegraphics[width=0.49\linewidth]{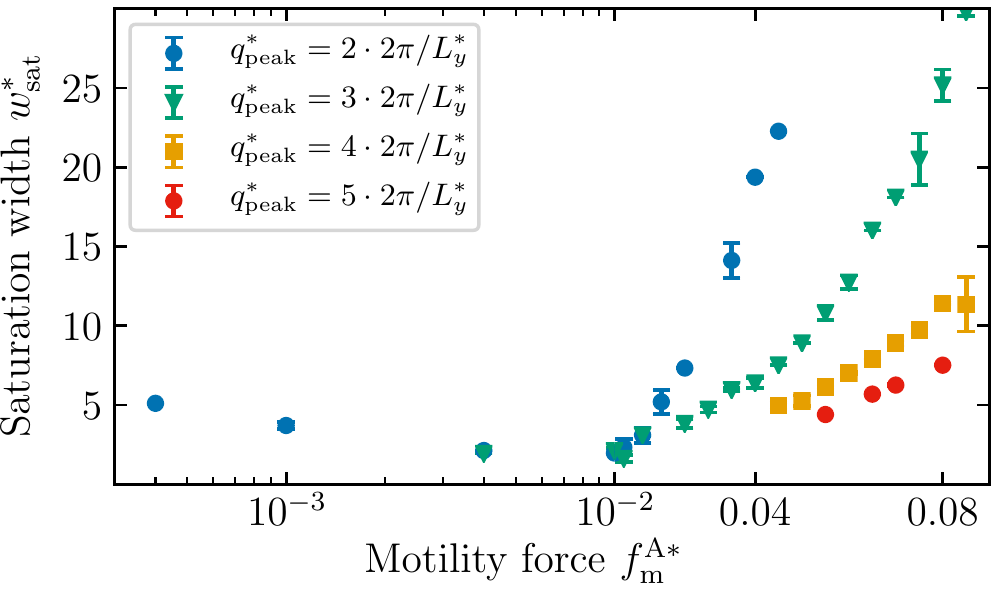}
	\includegraphics[width=0.49\linewidth]{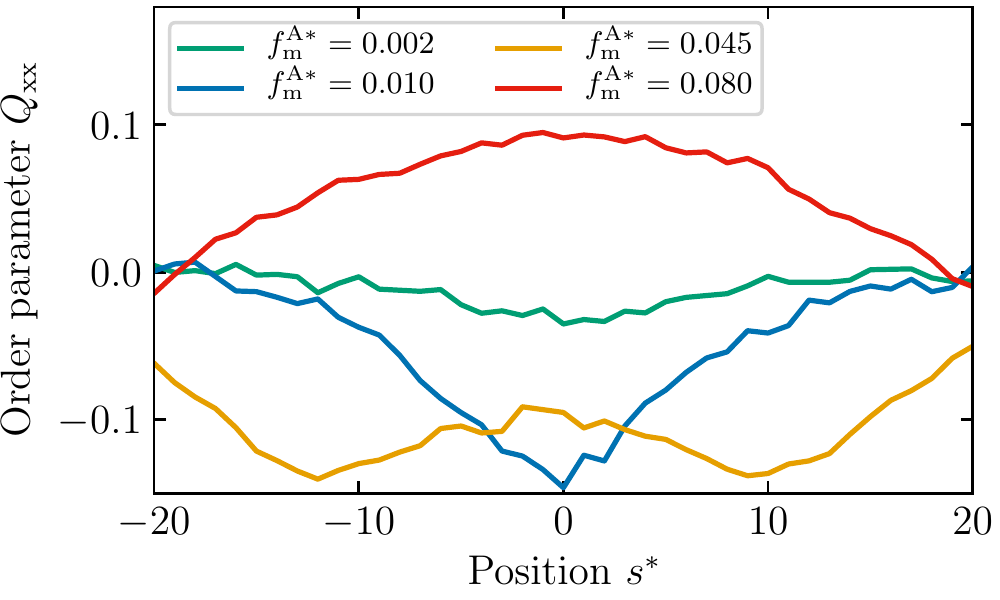}
	\caption{Interface velocity, interface (saturation) width, and order parameter dependence on motility-force strength of tissue~A in competitions with with a non-motile tissue. System size \mbox{$L_y^*=80$} in all, $\Delta\sigma_{\text{H}}^*=0$ in (b)-(d). (a) Interface velocity $v_{\text{int}}$  as a function of the motility-force strength $f_{\text{m}}^{\text{A}*}$ of tissue~A for various homeostatic stress differences $\Delta\sigma_{\text{H}}^* $. Dashed lines represent theoretical predictions according to Eq.~\eqref{int_velo}, with parameters fixed by independent simulations. Error bars display standard deviations (hidden behind markers). (b) Interface width $w^*$ as a function of time $t^*$ for different values of motility-force strength $f_{\text{m}}^{\text{A}\dagger}$ of tissue~A.
	 Note the logarithmic time scale, the interface width for non-vanishing motility is almost constant for 80\% of the time.
	(c) Saturation width $w^*_{\text{sat}}$ as a function of motility-force strength $f_{\text{m}}^{\text{A}*}$ of tissue~A for different peak wave vectors  $q_{\text{peak}}^*$.
	Note the logarithmic scale for $f_{\text{m}}^{\text{A}*}<0.01$. Error bars represent standard deviations. (d) Nematic order paramter $Q_{xx}$ as a function of the position $s^*$ for various  motility-force strengths $f_{\text{m}}^{\text{A}*}$ of tissue~A. Peak wave vector $q_{\text{peak}}^*=3\cdot 2\pi/L_y^*$ for all curves.}
	\label{bulk_motility}
\end{figure}

Figure \ref{bulk_motility}(a) displays a comparision between Eq. \eqref{int_velo}, with parameters fixed by independent simulations (see Refs.~\cite{Podewitz2015, Podewitz2016} for details), and the measured interface velocity, which shows very good agreement.
In the following, we focus on the proposed instabilities and study each of them individually.

\subsection{Bulk motility force}
We study first the effect of bulk motility without additional difference in homeostatic stress, i.e. the two tissues are identical except that tissue~A has a motility force $f_{\text{m}}^{\text{A}}>0$ while tissue~B is non-motile ($f_{\text{m}}^{\text{B}}=0$).  As predicted in Ref.~\cite{Williamson2018}, a prescribed motility force can drive interface propagation and the motile tissue invades the non-motile one at constant velocity. An instability is predicted for
\begin{equation}
\Delta v_{\text{f}} > \frac{2\Gamma(l_{\text{A}} \xi_{\text{A}} + l_{\text{B}} \xi_{\text{B}})}{l_{\text{A}} l_{\text{B}} \xi_{\text{A}}\xi_{\text{B}}(l_{\text{A}} +l_{\text{B}} )}\text{,}
\label{instability3}
\end{equation}
with difference in bulk velocity $\Delta v_{\text{f}} = f_{\text{m}}^{\text{A}}/\xi_{\text{A}} - f_{\text{m}}^{\text{B}}/\xi_{\text{B}}$ and interfacial tension $\Gamma$ \cite{Williamson2018}. 

\begin{figure}
	\centering
	\flushleft{\hspace{.3cm}(a)\hspace{9.69cm}(c)}\\
	\begin{tabular}{p{0.65\linewidth} p{0.35\linewidth}}
		\includegraphics[width=1.0\linewidth]{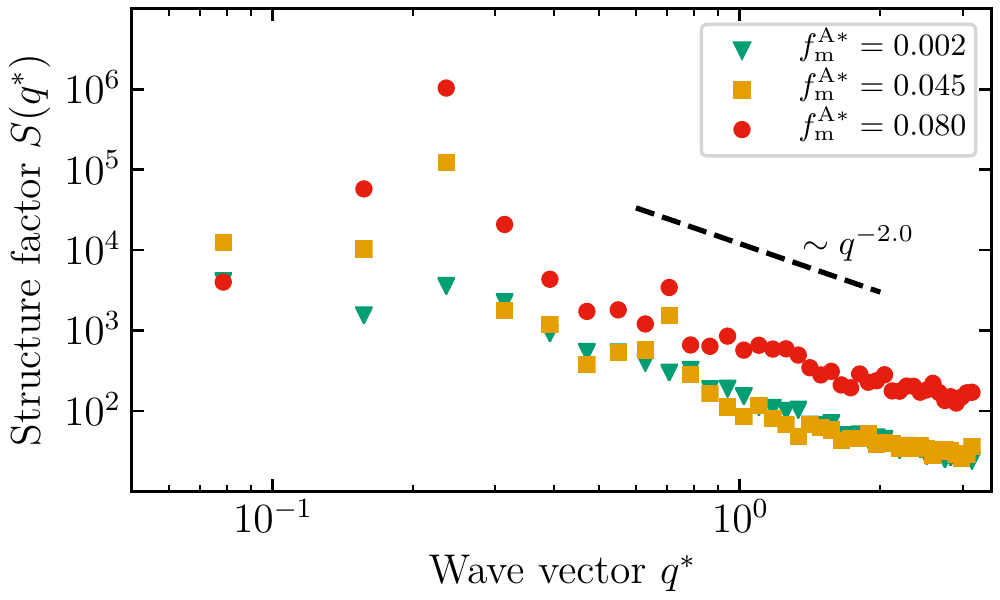}
		\vspace{-1.5cm}\flushleft{(b)}		\includegraphics[width=1.0\linewidth]{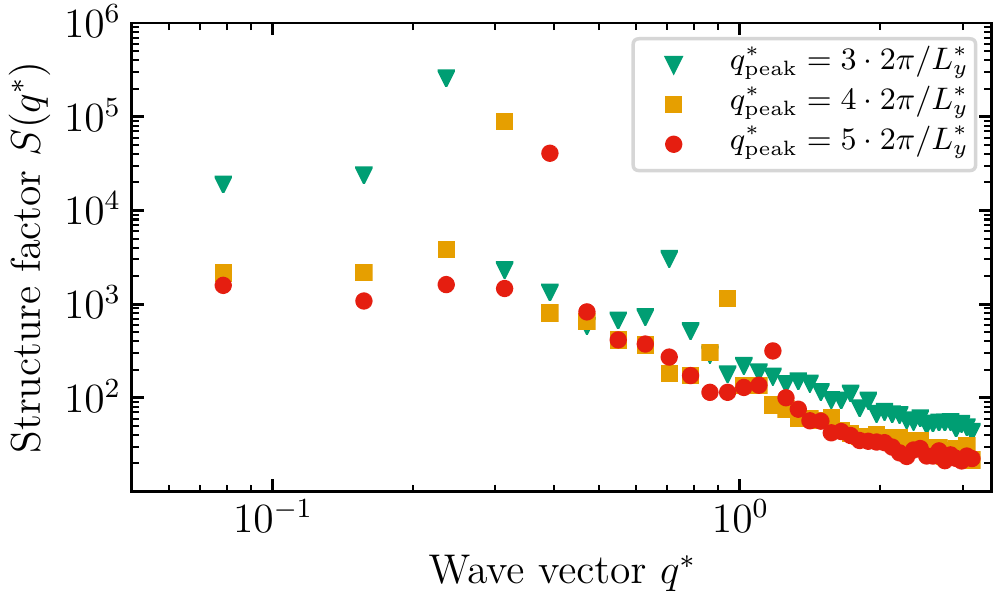} &
		\vspace{-6.25cm}	\includegraphics[width=1.0\linewidth]{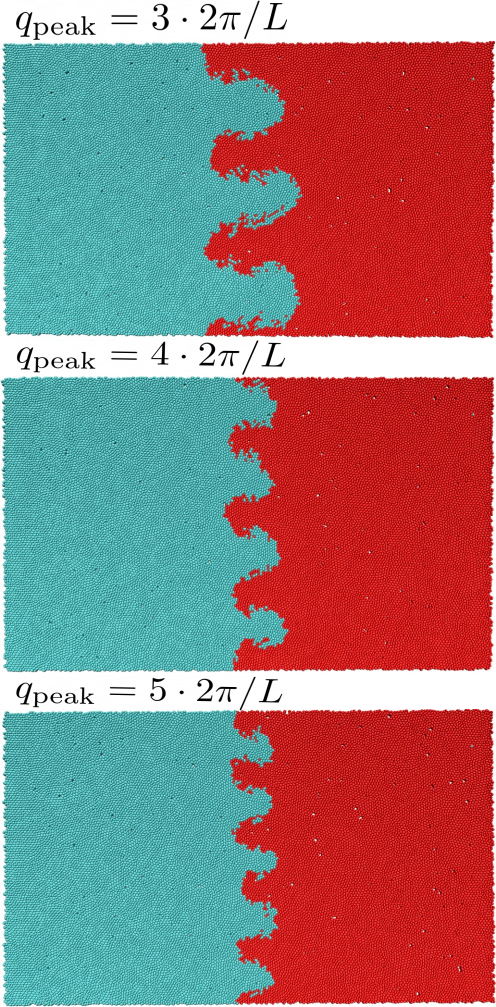}
	\end{tabular}
	\caption{(a) Structure factor $S(q^*)$ at the steady state for different values of the motility-force strength $f_{\text{m}}^{\text{A}*}$ of tissue~A for $\Delta\sigma_{\text{H}}^*=0$ and $q_{\text{peak}}^* = 3\cdot 2 \pi/L^*$. The dashed line is a guide to the eye. (b) Same as (a), but for fixed motility-force strength $f_{\text{m}}^{\text{A}*}=0.55$ and different peak wave vectors $q_{\text{peak}}^*$. (c) Snapshots obtained in the simulations of b) at the steady state at $t^*=80$. Note that the different stable peak wave vectors arise by chance from an initially flat interface. System size $L_y^*=80$ in all.}
	\label{structurefactor}
\end{figure}

Figure~\ref{snapshots} displays simulation snapshots for increasing motility-force strength of tissue~A. For vanishing motility force, the two competing tissues are identical, including the interaction between cells of different tissues, and thus the interface width $w(t) = \sqrt{\langle h^2 \rangle - \langle h \rangle^2}$ (with height field $h(y,t)$, see Ref.~\cite{Podewitz2016} for more details) diverges as a function of time (see snapshots in Fig.~\ref{snapshots} and blue line in Fig.~\ref{bulk_motility}(b), as well as Vid. S1 in the SI).
However, a rather small motility-force strength of tissue~A ($f_{\text{m}}^{\text{A}} \approx 5\cdot 10^{-4} G_0/R_{\text{PP}}^2$) suffices to arrive at a finite interface saturation width $w_{\text{sat}}$, i.e. small motility forces have a stabilizing effect on the interface (see snapshots in Fig.~\ref{snapshots} and green line in Fig.~\ref{bulk_motility}(b), as well as Vid. S2 in the SI). 
For larger motility-force strengths, protrusions of the motile into the non-motile tissue form at one particular finite wavelength. Over the time course of the first cell generation the interface width grows slowly with time ($w \sim t^{0.3}$). After the unstable wave mode has been selected,  the interface width increases linear with time ($w\sim t^{1.0}$). However, the mode amplitude does not grow indefinitely, but saturates at a motility-force dependent plateau due to nonlinear effects after about ten cell generations (see snapshots in Fig.~\ref{snapshots} and orange and red line in Fig.~\ref{bulk_motility}(b), as well as Vids. S3 and S4 in the SI).

The resulting wave pattern is remarkably stable over time once the steady state has been reached.
Figure~\ref{bulk_motility}(c) displays the saturation width $w_{\text{sat}}$ as a function of the motility-force strength. The saturation width first decreases with increasing motility-force strength, with $w_{\text{sat}}$ of the order of one or two cell layers at the minimum, i.e. an almost flat interface. For higher motility-force strength, the saturation width starts to increase and the aforementioned protrusions form, which we interpret as the onset of instability. Interestingly, independent simulations for identical parameter yield different wavelengths at the steady state. While the saturation width decreases with increasing $q_{\text{peak}}$ for identical $f_{\text{m}}^{\text{A}}$, the smallest motility-force strength at which a particluar wave mode is found increases with $q_{\text{peak}}$. This matches the predicted evolution of the most unstable wave mode in Ref.~\cite{Williamson2018}.

In order to study the observed interface patterns quantitatively, we calculate the time-averaged structure factor
\begin{equation}
S(q) = \langle \tilde{h}(q,t)\tilde{h}(-q,t)\rangle\text{,}
\end{equation}
at the steady state, where $\tilde{h}(q,t)$ denotes the spatial Fourier transform of the height field $h(y,t)$ (see Ref.~\cite{Podewitz2016} for further details). For self-affine surface growth the structure factor displays a power-law decay at the steady state \cite{Krug1997, Block2007}. Figure \ref{structurefactor}(a) shows the structure factor for the same values of motility-force strength as in Figs.~\ref{snapshots} and \ref{bulk_motility}(b). 
$S(q)$ displays deviations from a power-law decay by a peak at a certain wave vector larger than the system-spanning one (in Fig.~\ref{structurefactor}(a) $q_{\text{peak}}=3\cdot 2\pi/L$), which gets more pronounced for increasing motility-force strength and corresponds to the wavelength of the protrusions in Fig.~\ref{snapshots}. As mentioned above, for the same value of $f_{\text{m}}^{\text{A}}$, different wave vectors can become the dominating mode at the steady state in independent simulations. Figure~\ref{structurefactor}(b) displays the structure factor for three different peak modes for identical motility-force strength. The maximum decreases with increasing peak wave vectors, consistent with the higher saturation width for smaller $q_{\text{peak}}$ (see Fig.~\ref{bulk_motility}(c)).

 The stabilizing effect is accompanied by a preferred alignment of cells perpendicular to the interface, quantified by the nematic order paramter $Q_{xx} = 2p_x p_x-1$, with $p_x$ the $x$-component of the unit vector between the two cell particles. This leads to an active interfacial tension $\Gamma = \int_{-\infty}^{\infty} (\sigma_{yy} (s)- \sigma_{xx}(s)) \text{d}s$, due to cell growth \cite{Podewitz2016}.
Figure~\ref{bulk_motility}(d) displays the order paramter for different motility-force strengths. The overall alignment along the $y$-direction (i.e. negative $Q_{xx}$) first increases with growing $f_{\text{m}}^{\text{A}}$, with a maximum at the interface position. In the regime where protrusions start to form, the maximum splits into two maxima located to the left and the right of the interface, where the position of the maxima corresponds to the width of the protrusions. For even higher motility-force strength, when the saturation width becomes large, we find an overall alignment along the $x$-direction. 

\subsection{Homeostatic stress difference}

As shown in Refs.~\cite{Podewitz2016, Williamson2018}, interface propagation can be driven by homeostatic stress alone. For two tissues that only differ by their homeostatic stress, a stable interface propagating at constant velocity is found in the simulations \cite{Podewitz2016}. Two instability conditions for competition driven by a difference in homeostatic stress $\Delta\sigma_{\text{H}}$ have been proposed in Ref.~\cite{Williamson2018}, given by
\begin{align}
\Delta\sigma_{\text{H}} >& \frac{27}{4}\Gamma \frac{(\xi_{\text{A}}l_{\text{B}} - \xi_{\text{B}}l_{\text{A}})^2 (\xi_{\text{A}}l_{\text{A}} + \xi_{\text{B}}l_{\text{B}})}{l_{\text{A}}^2l_{\text{B}}^2(\xi_{\text{B}}-\xi_{\text{A}})^3}\text{, } \xi_{\text{B}} > \xi_{\text{A}}  \label{instability1}\\
\Delta\sigma_{\text{H}} >& 2\Gamma \frac{(\xi_{\text{A}}l_{\text{A}} + \xi_{\text{B}}l_{\text{B}})}{\kappa_{\text{B}}^{-1} - \kappa_{\text{A}}^{-1}} \text{, } \kappa_{\text{B}}^{-1} > \kappa_{\text{A}}^{-1} 
\label{instability2}
\end{align}

While substrate friction $\xi$ can be changed as an input parameter, the stress-response coefficient $\kappa$ is a tissue property, which needs to be determined in simulations and can not be controlled directly. In order to measure $\kappa$, we use a constant-stress ensemble and measure the growth rate as a function of the applied stress. $\kappa$ is then obtained by a linear fit \cite{Podewitz2015, Podewitz2016}. Since $\kappa=1/\chi\tau$, with the characteristic time $\tau$ for cell turnover, $\kappa$ can be changed by varying the apoptosis rate $k_{\text{a}}$. Reduction of $k_{\text{a}}$ yields a lower stress-response coefficient and thus a higher viscosity. 

\begin{figure}
	\centering
	\flushleft{\hspace{.cm}(a)\hspace{7.3cm}(b)}\\
	\includegraphics[width=0.49\linewidth]{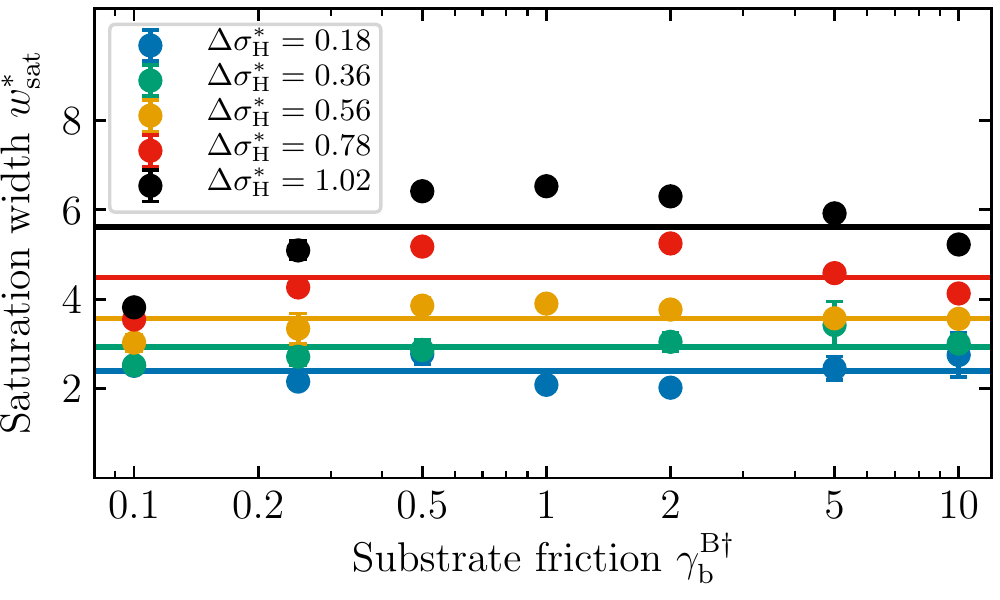}
	\includegraphics[width=0.49\linewidth]{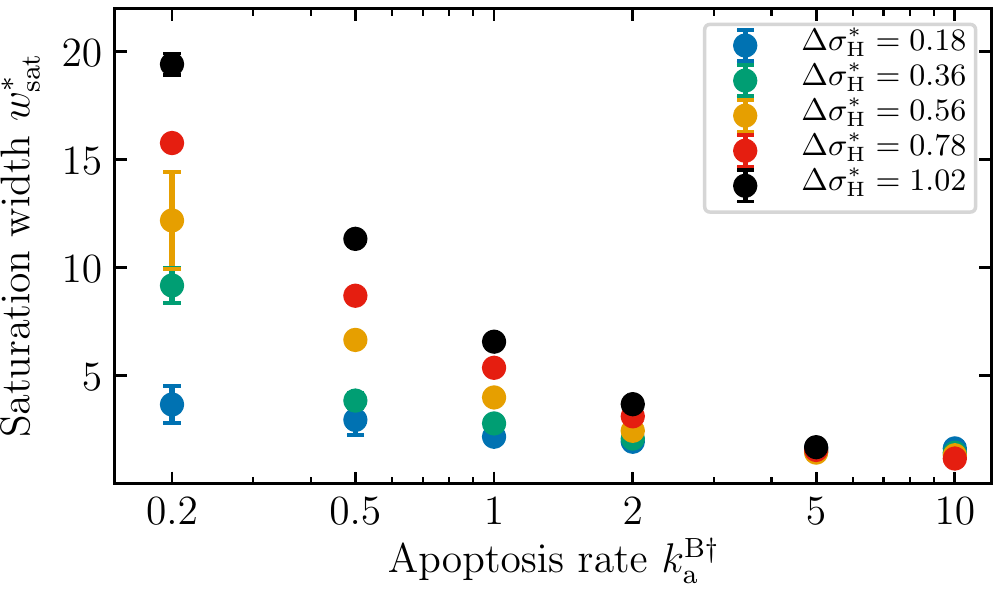}
	\caption{(a) Saturation width $w^*_{\text{sat}}$ as a function of substrate friction $\gamma_{\text{b}}^{\text{A}\dagger}$ of tissue~A for various homeostatic stress differences $\Delta\sigma_{\text{H}}^*$. (b) Same as (a) but as a function of the apoptosis rate $k_{\text{a}}^{\text{B}\dagger}$ of tissue~B. System size \mbox{$L_y^*=80$} in both.
		Error bars represent standard deviations. Note the different scales on the $y$-axis between (a) and (b).}
	\label{substrate}
\end{figure}

For different substrate frictions, we do not observe any instabilities, even for large differences in homeostatic stress. While the overall saturation width increases with growing homeostatic stress difference, $w_{\text{sat}}$ does not show systematic variations with substrate friction (see Fig.~\ref{substrate}(a)).

\begin{figure}
	
	\centering
	\vspace{-0.cm}
	\includegraphics[width=1.\linewidth]{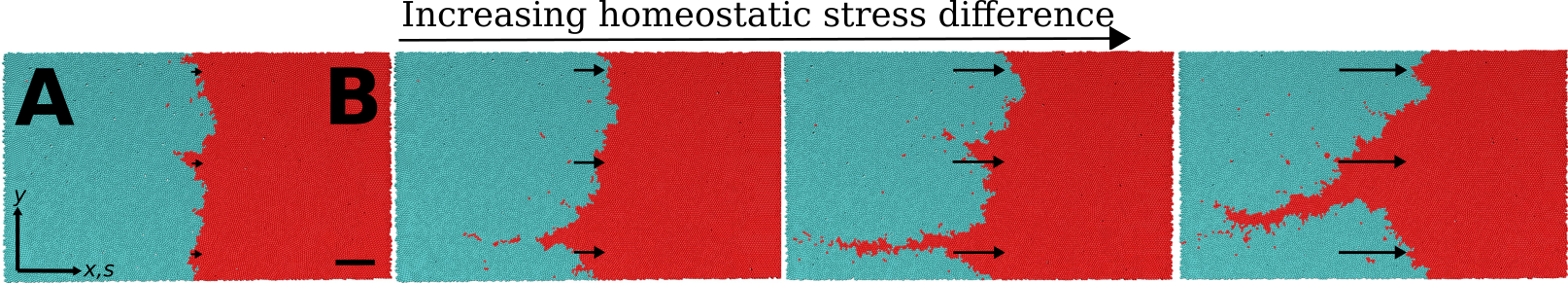}
	\caption{Simulation snapshots in competitions with different homeostatic stress differences $\Delta\sigma_{\text{H}}$ and reduced apoptosis rate $k_{a}^{\text{B}\dagger}=0.2$ of tissue~B. The interface moves towards tissue~B. From left to right: $\Delta\sigma_{\text{H}}=[0.18,\ 0.36,\ 0.56,\ 0.78]$. The length of the arrows corresponds to the distance the interface moves over five generations. System size $L_y^*=80$ and time $t^*=80$ in all.  The scale bar is 10 cell sizes. The snapshots are not represantative of a steady state, as fingers detach, disappear and reform over time.}
	\label{snapshots_apoptosis}
\end{figure}

According to Eq.~\eqref{instability2}, instabilities should only be obtained if the weaker tissue (the tissue with the higher homeostatic stress, here tissue~B) has a larger viscosity, i.e. a lower apoptosis rate than the stronger tissue. Figure~\ref{snapshots_apoptosis} displays simulation snapshots for different homeostatic stress differences. With increasing difference, a finger of the weaker tissue is found to develop into the stronger one. In contrast to the motility-driven case, no steady state is reached. The finger occasionally detaches, leaving a large island behind in the stronger tissue, moves along the interface and forms again (see Vid. S5 in the SI). However, we still find a mostly stable saturation width of the interface. Figure~\ref{substrate}(b) displays $w_{\text{sat}}$  as a function of the apoptosis rate $k_{\text{a}}^{\text{B}}$ of tissue~B for various different values of $\Delta\sigma_{\text{H}}$. We find that $w_{\text{sat}}$ increases for a reduced apoptosis rate (compared to $k_{\text{a}}^{\text{B}}=k_{\text{a}}^{\text{A}}$) above a critical homeostatic stress difference ($\Delta\sigma_{\text{H}}^{*}\approx0.4-0.5$), while the saturation width decreases for increased $k_{\text{a}}^{\text{B}}$, i.e. an enhanced apoptosis rate of the weaker tissue has a stabilizing effect. 

\begin{figure}
	\centering
	\flushleft{\hspace{.cm}(a)\hspace{7.3cm}(b)}\\
	\includegraphics[width=0.49\linewidth]{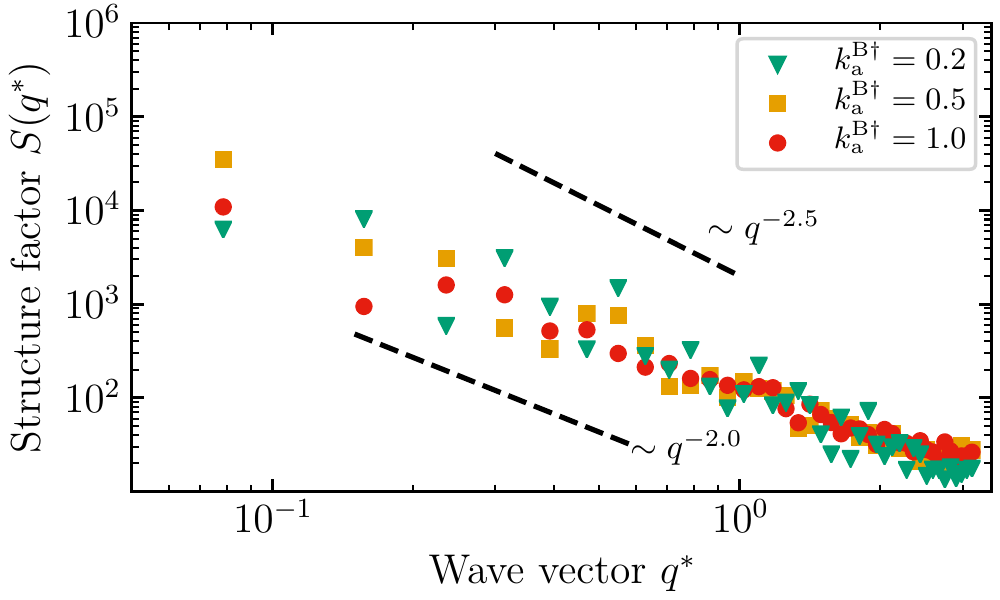}
	\includegraphics[width=0.49\linewidth]{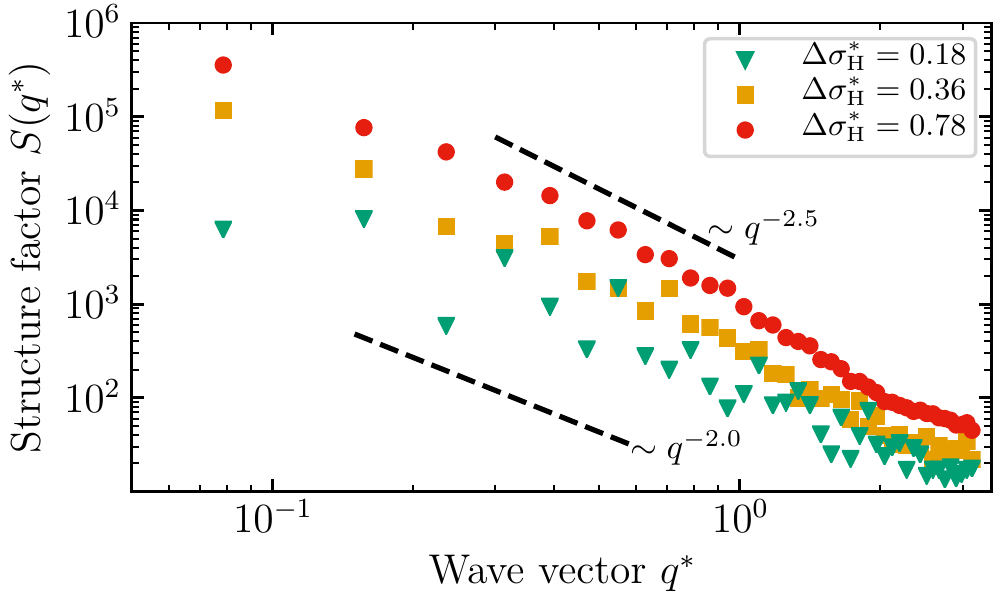} 
	\caption{(a) Structure factor $S(q^*)$ at the steady state for different values of the apoptosis rate $k_{\text{a}}^{\text{B}\dagger}$ of tissue~B for $\Delta\sigma_{\text{H}}^*=0.18$ (below the critical stress difference). The dashed lines are guides to the eye. 
		(b) Same as (a), but for different values of $\Delta\sigma_{\text{H}}^*$ and fixed $k_{\text{a}}^{\text{B}\dagger}=0.2$. System size $L_y^*=80$ in all.}
	\label{structurefactor_apoptosis}
\end{figure}

The structure factor reflects the increase in saturation width with growing homeostatic stress difference (see Fig.~\ref{structurefactor_apoptosis}). Below the critical stress difference, the structure factor for reduced apoptosis rate does not deviate significantly from the case of identical apoptosis rates of the competing tissues (see Fig.~\ref{structurefactor_apoptosis}(a)). However, for a fixed (reduced) apoptosis rate of tissue~B, the amplitude of all wave modes increases with growing  $\Delta\sigma_{\text{H}}$ (see Fig.~\ref{structurefactor_apoptosis}(b)), which matches the increase of the interface saturation width.  

\subsection{Bulk motility force \& homeostatic stress difference}

\begin{figure}
	\centering
	\flushleft{\hspace{.cm}(a)\hspace{8.9cm}(b)}\\
	\includegraphics[width=0.57\linewidth]{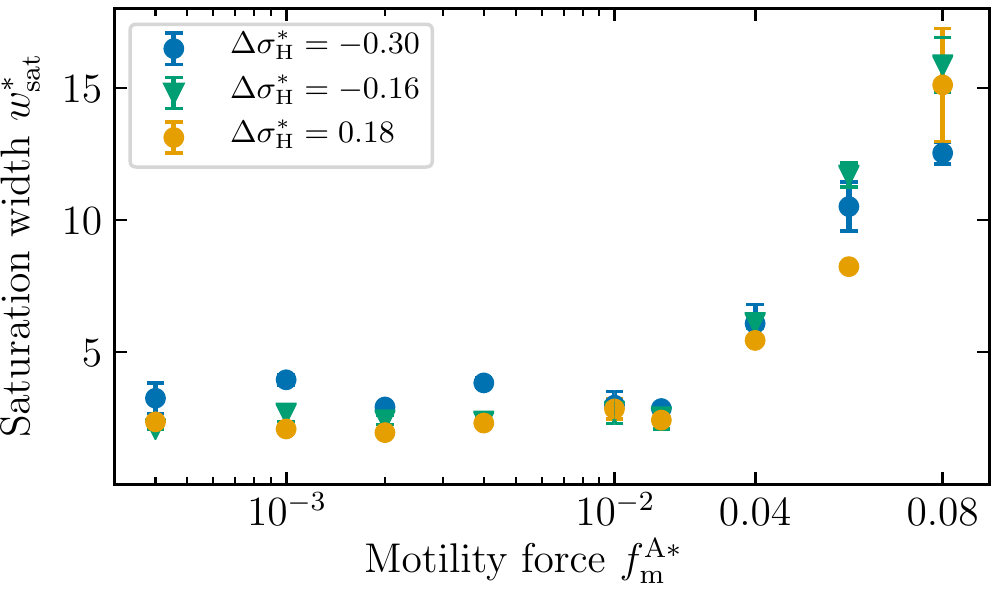}
	\includegraphics[width=0.42\linewidth]{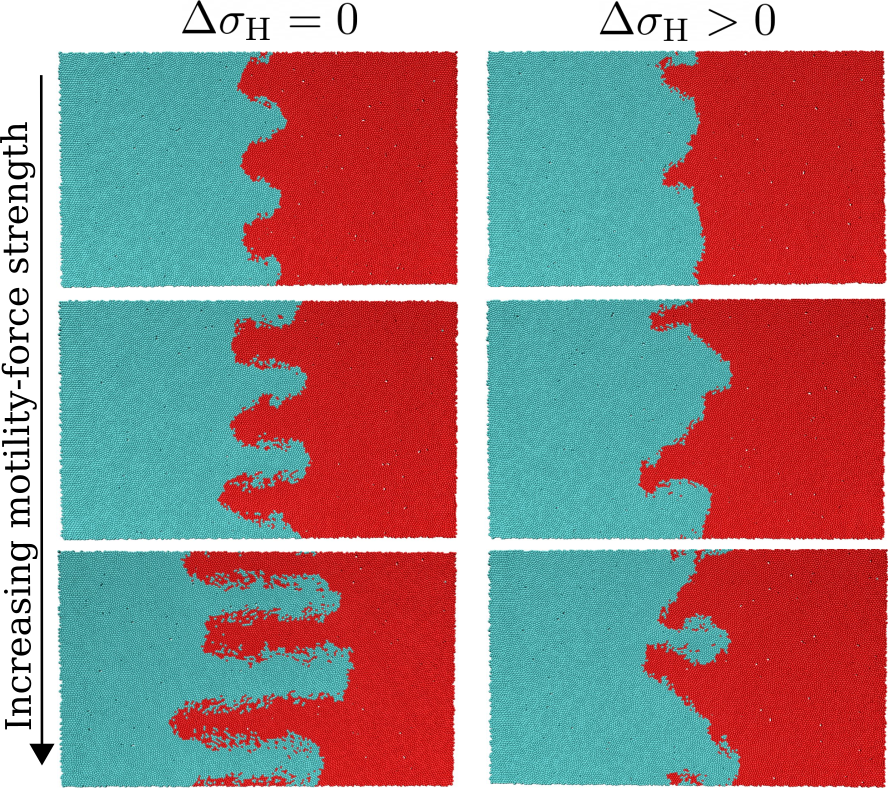}
	\caption{(a) Saturation width $w^*_{\text{sat}}$ as a function of the motility-force strength $f_{\text{m}}^{\text{A}*}$ of tissue~A for various homeostatic stress differences $\Delta\sigma_{\text{H}}^* $ in competitions with the standard tissue with $f_{\text{m}}^{\text{B}*} =0 $ and system size \mbox{$L^*=80$}. Note the logarithmic scale for $f_{\text{m}}^{\text{A}*}<0.01$. Error bars represent standard deviations. (b) Simulation snapshots for different motility-force strengts $f_{\text{m}}^{\text{A}*}$ of tissue~A, without (left) and with a homeostatic stress difference $\Delta\sigma_{\text{H}}^*=0.18$ (right). From top to bottom: $f_{\text{m}}^{\text{A}\dagger}=[0.04,0.06,0.08]$ The interface moves to the right. The tissues are otherwise identical. System size $L_y^*=80$ and time $t^*=80$ in all. Note that the snapshots with a homeostatic stress difference are not represantative of the steady state, as the interface shape is highly dynamic.}
	\label{combined}
\end{figure}

Finally, we take a closer look at a combination of differences in motility and homeostatic stress, with substrate friction and apoptosis rate identical for both tissues.
For small motility forces, the results of Ref. \cite{Podewitz2016} are not altered, the interface is stable and propagates at a constant velocity. In the regime where we find protrusions of the motile tissue into the non-motile tissue for vanishing homeostatic stress difference, the interface saturation width likewise starts to increase (see Fig.~\ref{combined}(a)). However, we do not observe protrusions at a particular wave length as for $\Delta\sigma_{\text{H}}^*=0$, but a highly dynamic shape of the interface (see snapshots in Fig.~\ref{combined}(b) for a comparision and Vid. S6 in the SI).

\section{Discussion}
We have investigated the stability of a propagating interface between two competing tissues over a broad parameter range in simulations of a particle-based model.

While the width of an interface between two tissues with identical properties diverges as a function of time, we find that already a very small directed bulk motility force of one tissue suffices to stabilize the interface at a finite width, similar to a  homeostatic stress difference \cite{Podewitz2016}. Above a critical motility-force strength, a single mode with wave length less than the system size becomes unstable. However, the amplitude of this mode does not diverge, as expected by linear-stability analysis, but nonlinear effects limit its growth, resulting in remarkably stable steady-state undulations of the interface. 
Cells align preferentially parallel to the interface for small motility forces, which transists into perpendicular alignment with growing motility-force strength. 

For interface propagation driven by a difference in homeostatic stress, an enhanced viscosity due to a reduced apoptosis rate of the weaker tissue results in an unstable interface above a critical homeostatic stress difference, reminiscent of a Saffman-Taylor instability.
The resulting pattern is much more dynamic than in the motility-driven case. A finger of the weaker tissue remains within the propagating front. This finger constantly reforms, moves and disappears. 

These two instabilities have recently been predicted by linear-stability analysis~\cite{Williamson2018}. For both instabilities, our results match the predicted evolution of the most unstable wave mode qualitatively. However, a quantitative comparision remains elusive, since we do not consider interfacial tension due to differential adhesion, as this would cause interfacial growth due to more free space for cells at the interface~\cite{Ganai2019}.

However, we do not observe the predicted instability for a difference in substrate friction of the competing tissues, even for large differences in the homeostatic stress between the competing tissues.

Our results suggest that interfacial patterns of competing tissues provide information about the underlying mechanical properties of the competing tissues. For example, a relatively regular --- almost sinusoidal  --- undulation pattern would suggest a motility-driven invasion, whereas a "remaining finger" of the host in the invading tissue would indicate a lower viscosity of the invader. However, experimental evidence of this kind of structures and instabilities will be needed before definite conclusions can be drawn.  From a theoretical perspective, possible future research directions on the stability of interfaces could be to account for anisotropic cell growth or enhanced interfacial growth rates~\cite{Ganai2019}.
\section{Acknowledgements}
The authors gratefully acknowledge the computing time granted through JARA-HPC on the supercomputer JURECA \cite{jureca} at Forschungszentrum J\"ulich. Parts of the simulations
were performed with computing resources granted by RWTH Aachen~University~under~project~rwth0475.
\section{Bibliography}
\bibliographystyle{iopart-num}
\bibliography{all.bib}
\end{document}